\documentclass[aps,notitlepage,twocolumn,a4paper,showpacs,preprintnumbers,amsmath,amssymb,floatfix]{revtex4}
\usepackage{amsmath,cleveref,times,graphicx,bm,nicefrac,tikz}

\newcommand{\e}{\text{e}}
\Crefname{figure}{Fig.}{Fig.}
\begin{document}
\title{Theoretical benchmarking of laser-accelerated ion fluxes by 2D-PIC simulations}

\author{F. Mackenroth}
\email{felix.mackenroth@chalmers.se}
\affiliation{Department of Physics, Chalmers University of Technology, SE-41296 G\"oteborg, Sweden}

\author{A. Gonoskov}
\affiliation{Department of Physics, Chalmers University of Technology, SE-41296 G\"oteborg, Sweden}
\affiliation{Institute of Applied Physics, Russian Academy of Sciences, Nizhny Novgorod 603950, Russia}
\affiliation{Lobachevsky State University of Nizhni Novgorod, Nizhny Novgorod 603950, Russia}

\author{M. Marklund}
\affiliation{Department of Physics, Chalmers University of Technology, SE-41296 G\"oteborg, Sweden}

\date{\today}
\begin{abstract}
There currently exists a number of different schemes for laser based ion acceleration in the literature. Some of these schemes are also partly overlapping, making a clear distinction between the schemes difficult in certain parameter regimes. Here, we provide a systematic numerical comparison between the following schemes and their analytical models: light-sail acceleration, Coulomb explosions, hole boring acceleration, and target normal sheath acceleration (TNSA). We study realistic laser parameters and various different target designs, each optimized for one of the acceleration schemes, respectively. As a means of comparing the schemes, we compute the ion current density generated at different laser powers, using two-dimensional particle-in-cell (PIC) simulations, and benchmark the particular analytical models for the corresponding schemes against the numerical results. Finally, we discuss the consequences for attaining high fluxes through the studied laser ion-acceleration schemes.
\end{abstract}

\pacs{41.75.Jv,52.38.Kd,52.59.-f,52.65.-y}
\maketitle
\section{Introduction}
Producing high-energy ions by compact, laser-based accelerators has received considerable interest and effort over the past decade \cite{Daido_etal_2012,Macchi_etal_2013,SSBulanov_etal_2008a}. The principal work horse of laser ion-acceleration up to now is the robust and widely studied scheme of target normal sheath acceleration (TNSA) \cite{Roth_etal_2002,Mora_2003,Cowan_etal_2004,Passoni_etal_2010}. This scheme, however, suffers from a number of severe drawbacks \cite{Tajima_etal_2009}, including (i) a disfavorable scaling of achievable ion energies with increasing laser intensity \cite{Robson_etal_2007}, limiting the scheme's efficiency at higher laser intensities, (ii) a limited control over the acceleration, as is apparent from the mostly thermal ion energy spectrum, hindering the precise tuning of the target ion energies, as well as (iii) a low efficiency of the acceleration mechanism at high laser powers \cite{Kim_etal_2013}. Many of these drawbacks, however, can be counteracted to some extent by specially designed targets \cite{Flippo_etal_2008,Buffechoux_etal_2010,Burza_etal_2011,Gaillard_etal_2011} and laser pulse shapes \cite{Markey_etal_2010,Pfotenhauer_etal_2010}. Thus, the comparatively simple design and operation of TNSA accelerators renders it a promising mechanism for the construction of reliable and stable lower-energy laser ion-accelerators. On the other hand, in order to compensate for the mentioned drawbacks, there have been a number of novel ion acceleration schemes proposed, such as Coulomb explosion (CE) of clusters \cite{Ditmire_etal_1997,Kovalev_etal_2007a,Kovalev_etal_2007b} and double-layered targets \cite{Esirkepov_etal_2002,Esirkepov_etal_2006,SSBulanov_etal_2008b}, designed to provide narrow energy spreads for the generated ion beams. Next to these schemes, which still rely on an expanding electron cloud due to local plasma heating, new schemes were introduced promising a more direct energy transfer from the laser pulse to an accelerated ion bunch, such as collisionless shock acceleration \cite{Silva_etal_2004,Haberberger_etal_2012}, hole boring (HB) \cite{Schlegel_etal_2009} or laser piston, also referred to as light sail (LS), \cite{Esirkepov_etal_2004,Bulanov_etal_2010,Henig_etal_2009,Kar_etal_2012}. The latter two schemes aim at directly employing the laser's light pressure to move the plasma electrons, which subsequently pull the heavier ions by the Coulomb field, thus promising a more controllable energy transfer and hence more easily tunable ion bunch properties as well as a higher efficiency. All the above mentioned laser ion-acceleration schemes surpass the capabilities of TNSA by aiming at higher laser intensities and powers, as will become available in the coming years \cite{Major_etal_2009,ExtremeLight}, whence one can subsume them under the general term \textit{high-power schemes}. 

The concept of TNSA has been well studied in experiment \cite{Roth_etal_2002} and is supported by a well developed theory \cite{Mora_2003}. Assessing the prospects of the mentioned high-intensity acceleration mechanisms is a still ongoing and demanding effort in both experiment and theory. Despite the analytical models developed for each mechanism, several issues make it difficult to directly compare their potentials. Firstly, each mechanism requires adjusting the parameters of the target, which can depend also on the laser pulse's parameters, such as total energy, duration, focal spot size or pulse contrast. Secondly, not only the peak or average energy, but also many other characteristics of the produced ions may be of crucial importance for certain potential applications. For instance, some schemes may provide higher ion energies, sacrificing the total number of accelerated particles, possibly even reducing the overall energy of the accelerated ion bunch. Apart from high ion energy, however, potential applications of laser-based ion sources impose requirements on such other characteristics of the ion beams, e.g., the particle number. Thus, the comparison of the high-intensity schemes' efficiencies needs to be performed carefully and thoroughly. Finally, there is only limited knowledge available about the agreement between the simplifying theoretical models and the complex physics of the ion acceleration over a broad range of parameters, as is necessary for a systematic overview.

In this work we aim at providing a systematic study of the respective efficiencies of the most common high-power laser ion-acceleration schemes and a comparison to numerical simulations. We include three of the high-power schemes mentioned above in this study, namely Coulomb explosion, hole boring and light sail, since they exemplify the fundamentally different physical processes and main driving mechanisms for laser-based acceleration schemes in solid targets. We assess the schemes' prospects using particle-in-cell (PIC) simulations and previously developed analytical models. To study these acceleration schemes we keep the laser parameters fixed and consider four classes of target designs, optimized for the various schemes (s. Fig.~\ref{fig:Targets}). This practice is motivated by striving to provide the most valuable information for laser facilities where the laser parameters are largely fixed but the target may be changed. For full comparability we analytically, as well as numerically, benchmark the results of the high-intensity schemes against TNSA, simulating a target optimized for TNSA as well. Finally, to provide a concise comparison of the studied laser acceleration schemes in terms of only one parameter, we use an ion momentum flux density as a basic efficiency measure applicable to different acceleration schemes.

The paper is organized as follows: After the introduction we provide a concise definition and discussion of the relativistic ion current, employed as the single parameter quantifying the laser-ion acceleration schemes' efficiency, employed laser parameters, the configuration of the performed simulations and an estimate of thermal noise level. Next, we summarize the known analytical theories and compare the theoretical predictions to a performance analysis in a two-dimensional PIC simulations for the three high-power acceleration schemes Coulomb explosion, hole boring and light sail. Finally, we compare all schemes to each other and benchmark their respective performance against TNSA.
\section{Analytical models and 2D PIC simulations}
\subsection{Relativistic ion current}
A possibly problematic issue in comparing the various different high-intensity ion acceleration schemes is that they have all been optimized and benchmarked with the emphasis put on different aspects of possible applications of high-energy ion beams. Thus, in each optimization different benchmark parameters were employed, such as peak energies \cite{Yin_etal_2006,Qiao_etal_2009,Bulanov_etal_2010}, number of accelerated particles \cite{Yu_etal_2005} as well as a small width of the ion distribution in energy \cite{Hegelich_etal_2006,SSBulanov_etal_2008b,Palmer_etal_2011,Haberberger_etal_2012,Ji_etal_2014} or space \cite{Chen_etal_2009}. In order to assess the performance and potential of the various ion acceleration schemes we are thus going to study a single parameter, in order to put all ion acceleration schemes in context. A suitable parameter for this task is the relativistic ion current, or flux of momentum, generated by the laser acceleration, defined as
\begin{align}
 \mathbb{J} = \int d^3r\ \gamma(\bm{r}) \frac{\bm{v}(\bm{r})}c n(\bm{r}),
\end{align}
where $\bm{r}=(x,y,z)$ is the spatial coordinate, $\bm{v}(\bm{r})$ is the ions' velocity, $n(\bm{r})$ the particle density, $\gamma(\bm{r}) = (1-(\bm{v}(\bm{r})/c)^2)^{-1/2}$ the Lorentz factor and the division by the speed of light $c$ turns the current into a dimensionless measure quantifying the total number of accelerated ions weighed with their relativistic factor. For a homogeneous ion bunch of constant velocity the current obviously reduces to the product of the particle number and the relativistic factor $\gamma \bm{v}/c$, indicating that the current indeed is a good measure for the bunch's overall energy content. We thus employ this parameter to benchmark all the studied ion acceleration schemes' efficiency in transforming the input laser energy into output ion beam energy. 

Since we are studying the acceleration schemes numerically in a two-dimensional geometry, however, we are going to study a two-dimensional equivalent of the ion current $\mathbb{J}$, labeled \textit{ion current density} $\bm{j}$. This takes into account that the transverse dimension, not resolved in our simulations, is assumed to feature translational symmetry. Thus, instead of the volumetric integral over the particle density $n(\bm{r})$ in our study we only perform a two-dimensional areal integral over the two dimensions resolved in the simulations
\begin{align}
 \bm{j} = \int dxdy\ \gamma(\bm{r}) \frac{\bm{v}(\bm{r})}c n(\bm{r}),
\end{align}
where we have chosen the target normal to coincide with the $x$-axis and labeled the resolved perpendicular coordinate $y$. Since we only consider the accessible ion current, propagating along the target normal direction the decisive, one-dimensional measure to quantify the acceleration schemes is
\begin{align}
 j_x \equiv j = \int dxdy \ \gamma(\bm{r}) \frac{v_x(\bm{r})}c n(\bm{r}).
\end{align}
Whereas the three-dimensional current is dimensionless, as argued above, in the present case the ion current density is given per unit distance in the non-resolved perpendicular coordinate direction. As a typical length unit of the studied ion acceleration we will always give it in $\left[j\right]=\mu\text{m}^{-1}$. In order to recover a conventional, dimensionless three-dimensional current one can multiply the simulation results with an assumed perpendicular extent of the accelerated ion bunch. As a good approximation this extent can be assumed to be equal to the ion bunch's diameter, which is of order of the laser spot size. 
\subsection{Laser parameters}
We aim at modeling laser parameters as are mostly available at present or upcoming high-power laser facilities. We thus consider a laser pulse of fixed wavelength $\lambda_0=810$ nm (frequency $\omega_0=2\pi c/\lambda_0$), FWHM duration $\tau_0=44$ fs, with a Gaussian profile focused to a FWHM beam waist radius $w_0=10\,\mu$m. We only vary the pulse's total energy $\varepsilon_0$. Its peak intensity is consequently given by $I=2\varepsilon_0\sqrt{\log2}/(\tau_0 w_0^2\sqrt{\pi})$. For the best performance of the mentioned high-power acceleration schemes we model the laser pulse to be circularly polarized and to hit the target under normal incidence, except in the simulation of TNSA, where a linearly polarized laser pulse hits the target at an incidence angle of $\pi/4$.

The targets are modeled with varying geometry but given number density $n_0 = 30 n_\text{cr}$ with the critical density $n_\text{cr}=m_e\omega_0^2/4\pi e^2$, where $m_e$ ($e < 0$) is an electron's mass (charge). The thickness of the target will be chosen for each acceleration scheme separately and is denoted by $d_\text{LS},d_\text{CE},d_\text{HB}$ for the respective schemes.
\subsection{Numerical configuration}
\begin{figure}[t]
 \includegraphics[width=\linewidth]{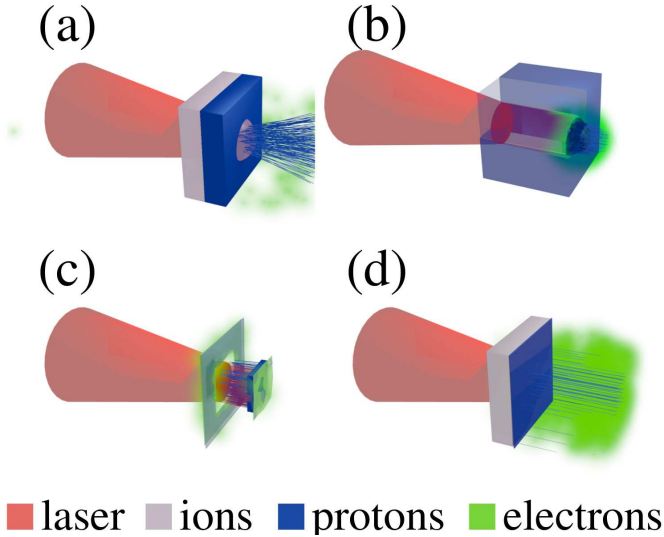}
 \caption{(color online) Laser (from left in red) impinging on optimized target designs for the studied high-power ion acceleration schemes: (a) for the study of Coulomb explosion we model a double-layered target consisting of electrons (green), protons (blue) and heavier ions (light purple), (b) for hole boring we consider a thick proton target, (c) for a light-sail we study a thin layer of protons and electrons and (d) for the benchmarking TNSA we study a thick target of heavy ions with a thin proton layer on the back.}
 \label{fig:Targets}
\end{figure}
In order to systematically benchmark the analytical models, it is required to compare their predictions to numerical solutions of the relevant plasma equations. We made use of the PIC code \textsc{Picador} \cite{Bastrakov_etal_2012}. We performed two-dimensional simulations, since they capture the one- or two-dimensional nature of the involved mechanisms as well as the possible formation of instabilities. Also, the benchmarking of the analytical models, all formulated in a one-dimensional geometry, against a two-dimensional simulation will already unveil whether significant drawbacks of its applicability are found.

We model targets consisting of electrons, protons (mass $m_p$, charge $-e$) and heavy ions (mass $20m_p$, charge $-e$), corresponding to the third ionization state of iron. In all studied acceleration schemes the heavy ions serve as a mount for the lighter protons and are not significantly accelerated. We thus focus the present work on the current of accelerated protons. In extracting the current density from the simulations consequently only protons were counted, weighed with their momentum along the target's normal along the $x$-direction. Generalising the results to other species of light ions is, however, straightforward.

While the analytical estimates are carried out irrespective of the specific target design, each of the calculations reviewed below assumes its underlying physical mechanism to work optimally. To ensure that this is indeed the case for each respective acceleration mechanism, one has to invoke certain, optimized targets. In order to fully assess the wide range of targets optimized for the various introduced high-power ion acceleration schemes we chose to perform numerical experiments on various different target geometries (s.~Fig.~\ref{fig:Targets}).

In the numerical spectra we only account for protons with kinetic energies above a certain lower cutoff energy, since experimentally there is only limited interest in the large abundance of low-energy, mostly thermal protons in the accelerated bunch. Furthermore, this exclusion reduces the impact of thermal noise that needs to be taken into account self-consistently in the simulations, as explained below. Thus, unless specifically mentioned, we employ a lower cutoff energy of $5$~MeV. For assessing the contribution of high-energy protons to the current, however, we also provide spectra for lower cutoff energies of $10$, $50$ and $75$~MeV.
\subsection{Thermal noise}
It is known that every solid target hit by a high-power laser is already heated by the laser's prepulse to a high temperature before the main pulse arrives. The interaction with the latter thus does not take place in a cold background plasma, but is surrounded by an abundance of hot ions and electrons in strong thermal motion. To account for this effect and to estimate the resulting background for the simulated proton currents, we initialize the particles in our numerical experiments with an initial temperature $T_0$. The corresponding random motion constitutes a current, due to thermal noise. To estimate the level of thermal noise, that could obscure the results of the simulations we estimate that each plasma particle in the simulation box has an initial thermal energy of $3k_BT_0/2$. Since we are only measuring the current in one spatial direction which, due to the equipartition theorem, comprises one third of the total thermal kinetic energy, the particles' nonrelativistic thermal momentum in the direction of the measured current will be given by
\begin{align}
 p_x^\text{th} = \sqrt{m_pk_BT_0}.
\end{align}
We fix the particles' initial temperature such that their kinetic energy is $k_BT_0 = 10^{-2}m_ec^2$. Thus, the protons' momentum along the current direction due to their initial thermal motion is given by 
\begin{align}
 p_x^\text{th} = \frac{c}{10}\sqrt{m_pm_e}.
\end{align}
The total current density caused by the thermal noise will then be
\begin{align}\label{Eq:Thermal_Current}
 j_\text{th} = \frac{p_x^\text{th}}{m_pc} \frac{N_0}{\Delta z},
\end{align}
where $N_0$ is the number of all particles in the simulation box and $\Delta z$ the box's extent in the perpendicular direction not resolved in the simulation. Given that up to the particle number all the above quantities are fixed in our simulations, we can provide an engineering formula for the thermal current in our simulations
\begin{align}
 j_\text{th} = \frac{1}{10}\sqrt{\frac{m_e}{m_p}} \frac{N_0}{\Delta z} \approx 2\times 10^{-3}\frac{N_0}{\Delta z}.
\end{align}
In all comparisons to numerical results we indicate this thermal noise level as a lower boundary below which the simulations' results need to be interpreted with great care.
\subsection{Acceleration by Coulomb explosion}
\subsubsection{Governing model}
The basic concept of laser-ion acceleration via Coulomb explosion relies on the Coulomb repulsion of residual ions, collectively stripped of a large portion of the corresponding electrons. As a specific example of this basic ion acceleration concept we are going to study a refined, double-layered setup of a purely Coulomb explosion target \cite{Esirkepov_etal_2002}: it was proposed to mount a thin layer of light ions, protons in this case, of thickness $d_\text{CE,l}$ on a layer of heavier ones of thickness $d_\text{CE}$. Once a laser pulse has ionized both layers and ejected all electrons, the heavier ions are going to stay in place, while the lighter ones are repelled by the Coulomb force onward, resulting ideally in a collimated, dense bunch of light ions. In order to estimate the scheme's efficiency, we have to find a reasonable assumption on the ratio of electrons expelled from the target by the laser pulse. As a strongly simplifying assumption we assume the target to be ionized only by the prepulse. Neglecting any shielding of the ions' charge as well as the proton layer, the ratio of electrons expelled from the layer of heavy ions can be roughly approximated as the ratio of the laser pulse's total energy $\varepsilon_0$ to the total number of electrons in the target $N_\text{CE} = n_0 d_\text{CE} \pi w_0^2$ and the energy required to remove one single electron from the heavy ion cloud $\Delta \varepsilon$. To estimate latter we employ the common assumption of the ions' potential to be one-dimensional up to a distance from the target of its transverse dimension $r_0 = 2 w_0$. The binding energy of an electron to the cloud of heavy ions then becomes $\Delta \varepsilon = 2\pi e^2 n_0 d_\text{CE} r_0$ and we find the ratio of electrons expelled from the heavy ion layer to be
\begin{align}
 \nu_\text{CE} = \frac{\varepsilon_0}{\Delta \varepsilon\, N_\text{CE}}.
\end{align}
In the case $\varepsilon_0>\Delta \varepsilon N_E$ we assume all the electrons to be expelled from the heavy ion layer. In the assumed one-dimensional geometry the electric field exerted by the remaining layer of heavy ions will then be \cite{Esirkepov_etal_2002}
\begin{align}
 E_\text{CE} = 2 \pi \nu_\text{CE} n_0 Z e d_\text{CE},
\end{align}
where $Z$ is the heavy ions' ionization level. One can then estimate that the light protons will be accelerated to energies of the order
\begin{align*}
 \varepsilon_\text{CE} = m_p c^2 + e E_\text{CE} r_0,
\end{align*}
where $r_0$ again is the acceleration length over which the the one-dimensional approximation is assumed to hold. The proton current density is then given by
\begin{align}
 j_\text{CE} = \frac{p_\text{CE}}{m_p} \frac{N_\text{CE}}{w_0} \frac{d_\text{CE,l}}{d_\text{CE}},
 \label{Eq_CE_Current}
\end{align}
where the proton momentum is given by $p_\text{CE} = \sqrt{\varepsilon_\text{CE}^2/c^2 - m_p^2c^2}$ and the additional factor $d_\text{CE,l}/d_\text{CE}$ results from the proton layer being thinner than the layer of heavy ions. 

\begin{figure}[t]
\includegraphics[width=\linewidth]{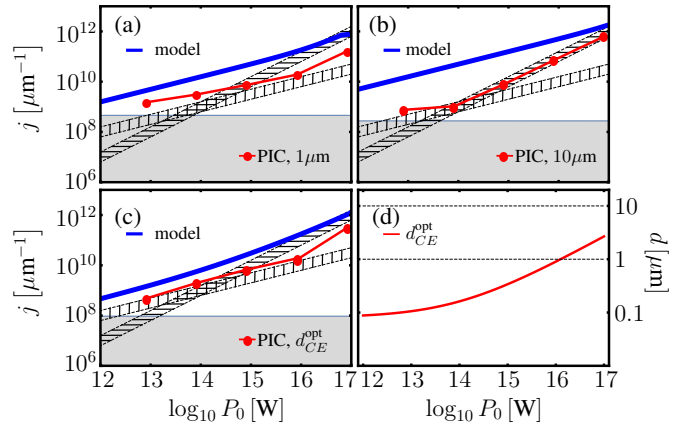}
\caption{(color online) Comparison of the proton currents resulting from three different target thicknesses in a Coulomb explosion setup for (a) $d_\text{CE}=1\,\mu$m, (b) $d_\text{CE}=10\,\mu$m and (c) $d_\text{CE}=d_\text{CE}^\text{opt}$ as compared to the theoretical model and (d) the dependence of the optimal target thickness on the laser power. For comparison [(a)-(c)]: linear scaling (horizontal stripes) and square root scaling (vertical stripes) with the laser power. Gray shaded area: thermal noise.}
\label{fig:CE_Currents}
\end{figure}
In order to provide an intuitive access to the proton currents to be expected according to the above theory, we derive a simple scaling law. Assuming the ratio of electrons expelled from the heavy ion layer does not saturate for the considered laser powers (as we confirmed numerically) we find the scaling $E_\text{CE}\sim\nu_\text{CE}\sim P_0$. The momentum of a proton accelerated in a Coulomb explosion setup can thus be estimated to be $p_\text{CE} = m_p c \sqrt{(\varepsilon_\text{CE}/m_pc^2)^2-1} \sim m_p \sqrt{E_\text{CE}/m_p} \sim P_0^{1/2}$, where $eE_\text{CE}r_0 \ll m_p c^2$ was used, as it holds for all studied laser powers. The resulting proton current density is then estimated to scale as
\begin{align}\label{Eq:CE_Current_scaling}
 j_\text{CE} \sim p_\text{CE} d_\text{CE,l} \sim \begin{cases}
                         \sqrt{P_0} & \text{fixed } d_\text{CE,l} \\
                         P_0 & d_\text{CE,l}^\text{opt}\sim\sqrt{P_0}. \\
                         \end{cases}
\end{align}
Finally, in order to run the Coulomb explosion mechanism optimally, it was found from simulations that the heavy ion layer should have an optimum thickness of \cite{Esirkepov_etal_2006}
\begin{align}\label{Eq:CE_dopt}
 d_\text{CE}^\text{opt} = (0.4 a_0 + 3) \frac{n_{cr}}{n_0} \lambda_0.
\end{align}
\subsubsection{Comparison with 2D PIC modeling}
%
We model the proposed double layer setup as consisting of a layer of protons mounted on a layer of heavy ions (s.~Fig.~\ref{fig:Targets} (a)). Keeping in mind that the proposed double-layer setup needs to be manufactured, however, we relax the complication of a very thin target by assuming the proton layer to be only half as thick (in contrast to the thickness ration $0.06$ as employed in \cite{Esirkepov_etal_2002}) as the heavy ion layer, which for the optimal thickness already has to be sub-$\mu$m thin (s.~Fig.~\ref{fig:CE_Currents} (d)). For comparing the acceleration scheme's performance for various target geometries, we study three different thicknesses of the heavy ion layer, namely the optimized, power-dependent thickness $d_\text{CE}^\text{opt}$ and two fixed thicknesses of $1 \mu$m or $10 \mu$m. The two fixed target thicknesses are chosen since $d_\text{CE}^\text{opt}$ is always smaller than $10\,\mu$m but grows larger than $1\,\mu$m for laser powers above $10^{15}$~W (s.~Fig.~\ref{fig:CE_Currents} (d)). 

We find that the exact solutions of Eq.~(10) for the targets of fixed thicknesses $1\,\mu$m or $10\,\mu$m indeed predicts a scaling $j_\text{CE}\sim P_0^{1/2}$ (s.~Fig.~\ref{fig:CE_Currents} (a),(b)), whereas for the target of optimal thickness $d_\text{CE}^\text{opt}$ the current's scaling is predicted to shift from $j_\text{CE}\sim P_0^{1/2}$ to $j_\text{CE}\sim P_0$ (s.~Fig.~\ref{fig:CE_Currents} (c)) in the same manner as the optimal thickness shifts from a constant value to a scaling $d_\text{CE}^\text{opt} \sim P_0^{1/2}$ (s.~Fig.~\ref{fig:CE_Currents} (d)).

\begin{figure}[t]
\includegraphics[width=\linewidth]{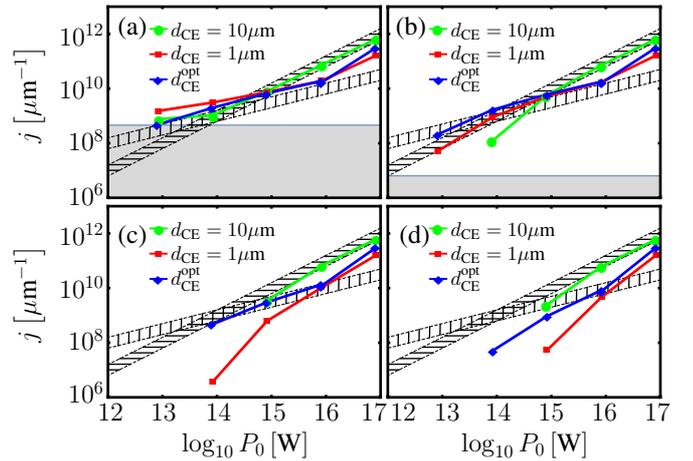} 
\caption{(color online) Direct comparison of all three thickness options for a Coulomb explosion setup above for a lower cutoff energy of (a) $5$ MeV, (b) $10$ MeV, (c) $50$ MeV and (d) $75$ MeV. For comparison: linear scaling (horizontal stripes) and square root scaling (vertical stripes) with the laser power. Gray shaded area: thermal noise.}
\label{fig:CE_Current_comp}
\end{figure}

Apparently the theoretically predicted proton currents are not reached in the performed PIC simulations, indicating the model overestimates the proton current by almost one order of magnitude. We associate this stark discrepancy with the theoretically neglected return currents, neutralizing the remaining heavy ions' Coulomb field as well as with an overestimation of the electron expulsion efficiency. Nevertheless, the performed simulations should still reproduce the found scaling laws and general tendencies. In the numerical simulations, on the other hand, only the $1\,\mu$m- and the $d_\text{CE}^\text{opt}$-targets approximately satisfy the predicted scaling $j_\text{CE}\sim P_0^{1/2}$ for low laser powers (s.~Fig.~\ref{fig:CE_Currents} (a),(c)). We note, however, that for the highest considered power $P_0 = 10^{17}$~W in the numerical simulation we find a significant portion of the laser pulse to break through the targets of optimal thickness and $d_\text{CE}=1\,\mu$m. Thus, the studied Coulomb explosion mechanism develops a substantial similarity to a light sail mechanism and the related data point exhibits a considerably higher proton current than one would expect from extrapolating the low-power scaling behavior (s.~Fig.~\ref{fig:CE_Currents} (a),(c)). The thicker target, on the other hand, exhibits a scaling $j_\text{CE}\sim P_0$ (s.~Fig.~\ref{fig:CE_Currents} (b)), which we attribute to a strong efficiency loss for lower laser powers since the electron return currents inside the thick target can efficiently compensate the heavy ions' Coulomb fields before a significant proton current can form. Due to this scaling of the proton current, the thickest target geometry produces a comparatively low overall current level for low laser powers, underrunning the predicted proton current by almost two orders of magnitude for $P_0=10^{14}$~W. The deviation from the linear scaling at a laser power $P_0=10^{13}$~W seems to be attributable to thermal noise, whence we do not include this laser power in the refined interpretation and consider in particular the $10\,\mu$m-target only down to a minimal laser power of $P_0=10^{14}$~W. We find the thickest target to produce the lowest proton current of all target geometries for this laser power (s.~Fig.~\ref{fig:CE_Current_comp} (a)). For laser powers $P_0\leq10^{15}$~W we find the intermediate thickness $1\, \mu$m to produce the highest proton currents, indicating a delicate balance between large numbers of accelerated particles and high proton energies (s.~Fig.~\ref{fig:CE_Current_comp} (a)). For laser powers $P_0\geq10^{15}$~W it seems that the target thickness, corresponding to the number of accelerated particles, dominates the proton current. This indicates that all target geometries produce comparable proton energies, with the apparent differences in the currents being attributable to different numbers of accelerated particles. This notion is also supported by comparing the proton currents resulting from the three different target geometries at different lower cutoff energies (s.~Fig.~\ref{fig:CE_Current_comp}). Including only protons above a lower energy cutoff of $50$~MeV in the current, the optimal thickness target yields a significantly larger current than the $1\,\mu$m thin target up to $P_0=10^{16}$~W, while the the $10\,\mu$m-target does not yield any current below $P_0=10^{15}$~W (s.~Fig.~\ref{fig:CE_Current_comp} (c)). This confirms that fixed thickness-targets at low laser powers produce proton currents of many, mostly low-energy thermal protons while the optimal, far sub-$\mu$m target thickness yields a true Coulomb explosion acceleration with fewer particles of higher energies. We can thus conclude that at low cutoff energies there is a strong thermal contribution to a Coulomb explosion ion acceleration setup. Only when excluding a large portion of low-energy protons does a non-thermal scaling $j_\text{CE}\sim P_0$ develop (s.~Fig.~\ref{fig:CE_Current_comp} (d)). At high laser powers, on the other hand, the currents do not change significantly relatively to each other, indicating that they all 
yield protons of comparable energies.
\subsection{Acceleration by hole boring}
\subsubsection{Governing model}
Unlike the previously discussed Coulomb explosion regime, hole boring denotes a radiation pressure acceleration in an overdense, thick target. The acceleration mechanism relies on the laser pulse piling up a thin electron spike, pushed forward by the ponderomotive force at the so-called piston velocity $v_\text{piston} = c\Xi/(1+\Xi)$, where the parameter $\Xi = \sqrt{I/(m_p n_0c^3)}$ \cite{Schlegel_etal_2009} is introduced. The resulting charge separation field pulls the ions behind, forming an electrostatic shock propagating through the target. We adopt a well-known analytical model predicting an ion velocity \cite{Schlegel_etal_2009}
\begin{align}
 v_\text{HB} = \frac{2 v_\text{piston}}{1+\left(\frac{v_\text{piston}}{c}\right)^2}.
\end{align}
The resulting proton momentum is then given by
\begin{align}
 p_\text{HB} = m_p c\frac{2 \Xi(1+\Xi)}{1+2\Xi}.
\end{align}
Consequently, the proton current density is given by
\begin{align}
 j_\text{HB} = \frac{p_\text{HB}}{m_p} \frac{N_\text{HB}}{w_0}.
\end{align}
In order to estimate the number of protons accelerated in the studied hole boring scheme, it is advantageous to note that the propagation speed of the shock front is given by $v_\text{HB} = c\Xi/(1 + \Xi)$. The time over which the laser pulse will be acting on the shock front can be approximated in the laboratory frame to be given by $t_\text{HB} = \tau_0/(1-v_\text{HB}/c)$. Approximating then the number of protons piled up in a hole boring setup after $t_\text{HB}$ by the total number of protons in the volume the shock front traversed, i.e., assuming perfect ionization and capture of the protons by the shock front, we find
\begin{align}
 N_\text{HB} = n_0 v_\text{HB} t_\text{HB} \pi w_0^2. 
\end{align}
Naturally, $N_\text{HB} < N_\text{HB}^\text{max} = n_0 d_\text{HB} \pi w_0^2$ has to always hold since the laser pulse cannot pile up more protons than are contained in its path through the whole thickness of the target. Thus, for a target that is too thin or an acceleration time which is too long, such that $ d_\text{HB} < v_\text{HB} t_\text{HB}$, the number of protons accelerated in hole boring is $N_\text{HB}^\text{max}$.

Finally, we provide an approximate scaling law for the proton current also for hole boring. For all studied laser powers it holds $\Xi \ll 1$ and from this parameter's definition one concludes $\Xi \sim P_0^{1/2}$. Thus we conclude $v_\text{HB}\sim p_\text{HB}\sim P_0^{1/2}$. The proton current density resulting from a hole boring setup thus is expected to scale as
\begin{align}\label{Eq:HB_Current_scaling}
 j_\text{HB} \sim p_\text{HB} N_\text{HB} \sim p_\text{HB} v_\text{HB} \sim P_0.
\end{align}
\subsubsection{Comparison with 2D PIC modeling}
Due to the conceptional simplicity of hole boring it is sufficient to only study one target geometry. We model a proton and electron layer three times thicker than one which allows for the onset relativistic transparency (Fig.~\ref{fig:Targets} b)). In this way it is ensured that the laser pulse does not break through the target and the HB regime is maintained over the whole acceleration process. Also, varying the target thickness was found to have no effect on the proton current.
\begin{figure}[t]
\includegraphics[width=\linewidth]{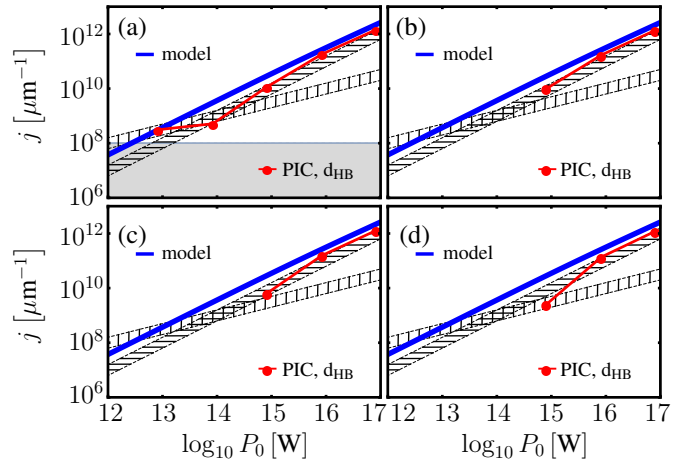} 
\caption{(color online) Comparison of the proton current resulting from a hole boring setup as compared to the model for a lower cutoff energy of (a) $5$ MeV, (b) $10$ MeV, (c) $50$ MeV and (d) $75$ MeV. For comparison: linear scaling (horizontal stripes) and square root scaling (vertical stripes) with the laser power. Gray shaded area: thermal noise.}
\label{fig:HB_comp}
\end{figure}
As apparent from the theoretical considerations presented above, we assumed a full ionization of the target as well as the accelerated current to be formed by all protons in the laser pulse's path. From the good agreement between the presented theory and PIC simulations accounting for all protons above a lower cutoff energy of $5$~MeV (s.~Fig.~\ref{fig:HB_comp} (a)) we conclude the former assumption to be reasonable and the presented theory to describe the physical acceleration process satisfactorily. Furthermore, we find the predicted linear scaling with the laser power to be very well reproduced and even slightly overrun. Only at the lowest considered laser power $P_0 = 10^{13}$~W the PIC simulation deviates from a linear scaling, which we attribute to the thermal noise level. From comparing the proton currents at various lower cutoff energies, where the absence of a data point indicates that at the given laser power there is no current of protons above the chosen cutoff energy formed (s.~Fig.~\ref{fig:HB_comp} (b)-(d)), we infer that at low laser powers hole boring yields a current of a large number of rather low-energetic particles. At large laser powers, however, hole boring reliably produces a large number of high energy protons, as we infer from the fact that the proton current at laser powers $P_0 \geq 10^{15}$~W do not significantly differ for a lower cutoff energy of $50$~Mev (s.~Fig.~\ref{fig:HB_comp} (c)) or $75$~Mev (s.~Fig.~\ref{fig:HB_comp} (d)).
\subsection{Light sail acceleration}
\subsubsection{Governing model}
Just like hole boring, the laser piston, or light sail, regime relies on directly employing the longitudinal radiation pressure for ion acceleration. We adopt the theoretical description of \cite{Esirkepov_etal_2004} which is somewhat more involved than the previously studied static acceleration mechanisms of Coulomb explosion and hole boring. The proton dynamics in a light sail regime are governed by the set of coupled differential equations \cite{Daido_etal_2012,Macchi_etal_2013}
\begin{align}
\begin{array}{r c l}
\displaystyle \frac{dp}{dt} &=& R(p) \frac{\displaystyle E^2_0(t-x/c)}{\displaystyle 2 \pi n_0 d_\text{LS}} \frac{\displaystyle \sqrt{p^2 + m_p^2c^2} - p}{\displaystyle \sqrt{p^2 + m_p^2c^2} + p}\\
p &=& m_p \,v \displaystyle \sqrt{\frac{\displaystyle 1}{\displaystyle 1-\left(\frac{v}{c}\right)^2}} \\
\displaystyle \frac {dx}{dt} &=& v.
\end{array}
\end{align}
The momentum dependent reflectivity is given by \cite{Macchi_etal_2009}
\begin{align}
 R(p) &= \begin{cases}
	  \frac{\pi \sigma^2(p)}{1 + (\pi \sigma(p))^2} & \text{if }\ a_0 < \sqrt{1 + \pi^2 \sigma^2(p)}\\
	  \frac{\left(\pi \sigma(p)\right)^2}{1 + a_0^2} & \text{if }\ a_0 > \sqrt{1 + \pi^2 \sigma^2(p)}
     \end{cases},
\end{align}
where the boosted areal density of the light sail in the comoving reference frame is given by $\sigma(p)=\sigma_0 \lambda(p)/\lambda_0$, with the momentum dependent wavelength in the light sail's moving frame $\lambda(p)$ and the dimensionless areal density in the laboratory frame $\sigma_0= n_0 d_\text{LS}/n_\text{cr} \lambda_0$. It was shown, that in an ideal light sail configuration with constant reflectivity $R\equiv1$ and constant laser field strength $E_0=\text{const}.$, a proton sheath acquires a momentum \cite{Esirkepov_etal_2004}
\begin{align}
 p_\text{LS} = m_p c \left(\text{sinh}\left(u_\text{LS}\right) - \frac{\text{csch}\left(u_\text{LS}\right)}{4}\right),
\end{align}
where we use $u_\text{LS}=\text{asinh}\left(\Omega_\text{LS} t_\text{LS} + h_\text{LS}^{3/2} + 3/2 h_\text{LS}\right)/3$, $\Omega_\text{LS} = (3 E_0^2)/(2 \pi n_0 d_\text{LS} m_pc)$ and $h_\text{LS} = p_{LS,0}/m_pc + \sqrt{1 + p_{LS,0}^2/m_p^2c^2}$. The time over which the light sail acceleration is expected to be maintained was estimated to be $t_\text{LS} = 2 \tau_0 (\varepsilon_0/(n_0 \pi w_0^2 d_\text{LS} m_p c^2))^2 /3$ \cite{Esirkepov_etal_2004}. Naturally, $\tau_0$ is a lower limit for $t_\text{LS}$, so for $(\varepsilon_0/(n_0 \pi w_0^2 d_\text{LS} m_p c^2))^2 <3/2$ we set $t_\text{LS} = \tau_0$. As we will see later, the differences in final light sail momenta between the exact and the approximate theory are marginal for larger laser powers, proving approximating the laser pulse to be of constant amplitude and the light sail perfectly reflecting to be admissible in this parameter regime. The proton current density is then
\begin{align}
 j_\text{LS} = \frac{p_\text{LS}}{m_p}\frac{N_\text{LS}}{w_0}.
\end{align}
The number of protons accelerated in an ideal light sail setup is simply given by all particles contained in the sheath thickness $N_\text{LS}=d_\text{LS}n_0 \pi w_0^2$.

In order to run the light sail mechanism optimally, it was shown that the target should have a thickness, balancing between reflection of the laser pulse and a premature break-through, due to the onset of relativistic transparency \cite{Macchi_etal_2013}. The optimum thickness was found to be given by \cite{Macchi_etal_2009}
\begin{align}\label{Eq:LS_dopt}
 d_\text{LS}^\text{opt} = \frac{a_0}{\pi} \frac{n_{cr}}{n_0} \lambda_0.
\end{align}
We finally estimate the scaling law for the proton current expected from a light sail setup. Inserting the definition of the parameters we derive for the optimal thickness $d_\text{LS}^\text{opt}$ the following scaling $\Omega_\text{LS}t_\text{LS} \sim P_0^{3/2}$ for $(\varepsilon_0/(n_0 \pi w_0^2 d_\text{LS} m_p c^2))^2 > 3/2$ and $\Omega_\text{LS}t_\text{LS} \sim P_0^{1/2}$ for $(\varepsilon_0/(n_0 \pi w_0^2 d_\text{LS} m_p c^2))^2 < 3/2$, where we respected the varying scaling of $t_\text{LS}$. Furthermore, the balance condition $(\varepsilon_0/(n_0 \pi w_0^2 d_\text{LS} m_p c^2))^2 = 3/2$ signifies the balance point $\Omega_\text{LS}t_\text{LS} \approx 1$, whence for $(\varepsilon_0/(n_0 \pi w_0^2 d_\text{LS} m_p c^2))^2 < 3/2$ we can approximate $\text{sinh}(u_\text{LS}) \sim \Omega_\text{LS}t_\text{LS}$ and for $(\varepsilon_0/(n_0 \pi w_0^2 d_\text{LS} m_p c^2))^2 > 3/2$ we find $\text{sinh}(u_\text{LS}) \sim (\Omega_\text{LS}t_\text{LS})^{1/3}$. In both cases the proton momentum thus scales as $p_\text{LS} \sim P_0^{1/2}$ and for the proton current density we expect a scaling
\begin{align}\label{Eq:LS_Current_scaling}
 j_\text{LS} \sim p_\text{LS} d_\text{LS}^\text{opt} \sim P_0.
\end{align}
A similar argument shows that for a fixed target thickness one expects the same linear scaling.
\subsubsection{Comparison with 2D PIC modeling}
\begin{figure}[t]
\includegraphics[width=\linewidth]{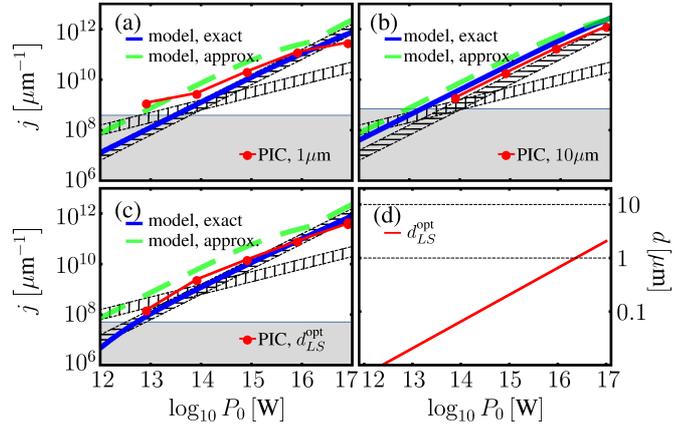}
\caption{(color online) Comparison of the proton currents resulting from a light sail setup for three different target thicknesses (a) $d_\text{LS}=1\,\mu$m, (b) $d_\text{LS}=10\,\mu$m and (c) $d_\text{LS}=d_\text{LS}^\text{opt}$ as compared to the exact and approximate theoretical models. (d): dependence of the optimal target thickness on the laser power. For comparison [(a)-(c)]: linear scaling (horizontal stripes) and square root scaling (vertical stripes) with the laser power. Gray shaded area: thermal noise.}
\label{fig:LS_comp}
\end{figure}
As a simple target is conceptionally required for light sail acceleration, we study a thin layer of protons and electrons (Fig.~\ref{fig:Targets} c)). Analogous to the Coulomb explosion regime, in order to study the acceleration's efficiency for varying target geometries we investigate three different layer thicknesses, namely the optimized, power-dependent thickness $d_\text{LS}^\text{opt}$ and two fixed thicknesses of $1 \mu$m and $10 \mu$m. We find the exact and approximate theoretical models introduced above to agree good within an order of magnitude (s.~Fig.~\ref{fig:LS_comp} (a)-(c)). The overestimation of the proton current by the approximate model at low laser powers is due to assuming a perfect reflectivity of the relativistic mirror, that is not yet fully achieved at the corresponding low radiation pressures. Also the PIC simulations follow the theoretical predictions in scaling as well to a very good extent, also in absolute numbers. There is, however, a slight deviation from the scaling $j_\text{LS}\sim P_0$ for the optimal thickness $d_\text{LS}^\text{opt}$ (s.~Fig.~\ref{fig:LS_comp} (c)). This behavior can most probably be attributed to imperfect proton capture due to the strong transversal ponderomotive force pushing the protons out of the interaction region and other onsetting plasma instabilities driven by high power lasers. A more discontinuous change in the scaling behavior is observed for the thin target $d_\text{LS}=1\,\mu$m at the highest laser power $P_0=10^{17}$~W (s.~Fig.~\ref{fig:LS_comp} (a)). This abrupt efficiency loss is due to the target's thickness falling below the optimal target thickness (s.~Fig.~\ref{fig:LS_comp} (d)) and thus containing too few protons to follow the linear trend. A similar, particle number-dominated explanation most likely applies to the behavior of the over-thick light sail target $d_\text{LS}=10\,\mu$m which maintains the linear scaling $d_\text{LS}^\text{opt}\sim P_0$ over all studied laser powers. Since the target, however, is always too thick to admit relativistic transparency and thus break-through of the laser pulse through the target, this target rather operates in a hole-boring regime, as is also supported from the close similarity of the simulations results to those of hole boring (s.~Fig.~\ref{fig:HB_comp} (a)). We also see that light sail operation at low laser powers below $10^{14}$~W requires sub-$\mu$m thin targets and thus ultra-high laser contrast.
\begin{figure}[t]
\includegraphics[width=\linewidth]{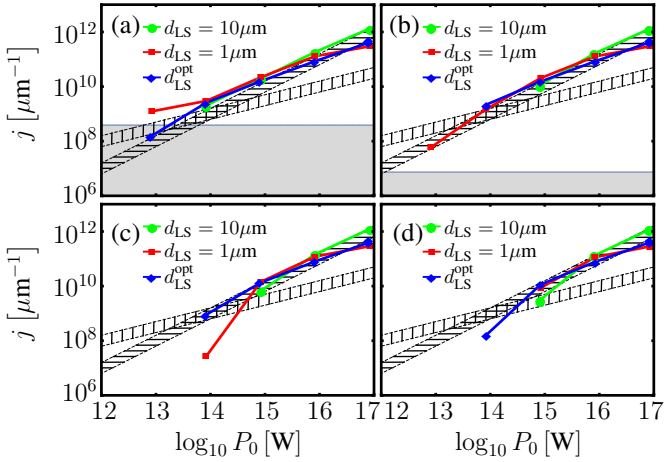}
\caption{(color online) Direct comparison of all three thickness options for a light sail setup above for a lower cutoff energy of (a) $5$ MeV, (b) $10$ MeV, (c) $50$ MeV and (d) $75$ MeV. For comparison: linear scaling (horizontal stripes) and square root scaling (vertical stripes) with the laser power. Gray shaded area: thermal noise.}
\label{fig:LS_comp_Cutoffs}
\end{figure}
Furthermore, it shows that at a laser power $P_0=10^{13}$~W the $1\,\mu$m thin target yields larger proton currents than the target with optimal thickness (s.~Fig.~\ref{fig:LS_comp_Cutoffs} (a)), merely because it yields larger particle numbers. When considering only protons above an energy of $10$~MeV this effect is even more pronounced as the optimal thickness target then does not yield any proton current, while for the $1\,\mu$m thin target there are a few, mostly thermal protons reaching this energy (s.~Fig.~\ref{fig:LS_comp_Cutoffs} (b)). For a laser power $P_0=10^{14}$~W, on the other hand the target of optimal thickness produces a current of particles almost exclusively at energies larger than $50$~MeV (compare Fig.~\ref{fig:LS_comp_Cutoffs} (b) and (c)), while the $1\,\mu$m-target yields a current of apparently more particles, however at lower energies (s.~Fig.~\ref{fig:LS_comp_Cutoffs} (c)). We thus conclude that at such smaller laser powers only the optimal, far sub-$\mu$m target thickness yields a true light sail acceleration, while the $1\,\mu$m-target produces a rather thermal proton current. Lastly, at the largest laser power $P_0=10^{17}$~W we find the currents from the target with optimal thickness to surpass those from the $1\,\mu$m-target for all lower cutoff energies, due to the latter's efficiency drop at these high laser powers (s.~Fig.~\ref{fig:LS_comp_Cutoffs} (a)). This behavior prevails up to lower cutoff-energies of $75$~MeV. The proton currents from the light sail target with $d_\text{LS}=10\,\mu$m are not particularly discussed here, since they need to be interpreted with great care due to the hole boring mechanism that comes additionally into play.
\subsection{Target normal sheath acceleration}
In addition, we consider a target design optimized for TNSA, in order to benchmark the high-energy schemes against the most conventional thermal acceleration scheme. To this end, we model a linearly polarized laser pulse to hit a typical TNSA target with thickness of $3\mu$m with a $0.2\mu$m thin hydrogen layer on the back surface (s.~Fig.~\ref{fig:Targets} (d)) under an incidence angle of $\pi/4$. We then compare the efficiency of the three mentioned high-intensity acceleration schemes to the performance of conventional TNSA.
\subsubsection{Governing model}
The modeling of this acceleration mechanism is done based on the well-known, one-dimensional Mora model of a heated plasma expanding into vacuum \cite{Mora_2003} along the $x$-direction with a velocity distribution $v_\text{TNSA} = c_s + x/t$ and a density profile $n = n_0 \text{exp}\left[-\left(1+x/(c_st)\right)\right]$, where $c_s = \sqrt{(Z k_B T)/m_p}$ is the speed of sound, $Z$ the ions' ionization level, $k_B$ the Boltzmann constant and $T$ the plasma temperature. The model is only defined for $x>x^\text{min}(t)=-c_st$ and the front of the expanding ion cloud is located at $x^\text{max}(t)=(2 \log [\omega_{p,i} t] - 1) c_st$. Consequently, the ion current density, given by the product of the transverse size of the acceleration region, approximated by $w_0$ with the spatial integral over the product of the particle density and the respective velocity through the plasma expansion volume, is given by

\begin{align}
j_\text{TNSA} &= n_0 w_0 \int_{x^\text{min}(t_\text{TNSA})}^{x^\text{max}(t_\text{TNSA})} dx \sqrt{\frac{v_\text{TNSA}^2}{1-\left(\frac{v_\text{TNSA}}{c}\right)^2}} \e^{-\left(1+\frac{x}{r_\text{TNSA}}\right)},
\end{align}
where we defined the distance a plasma perturbation travels during the operation time of TNSA $r_\text{TNSA}=c_s t_\text{TNSA}$ with $t_\text{TNSA} = 1.3 \tau_0$ an effective estimate of how long the TNSA mechanism can be upheld and the additional square root factor is the position dependent Lorentz factor of the proton cloud. The proton plasma frequency is given by $\omega_{p,i} = \sqrt{(4 \pi n_0 e^2)/m_p}$ and as an estimate for the plasma temperature we employ the ponderomotive model $T = m_e \left(\sqrt{1 + a_0^2} - 1\right)$.
\subsubsection{Comparison with 2D PIC modeling}
\begin{figure}[t]
\includegraphics[width=\linewidth]{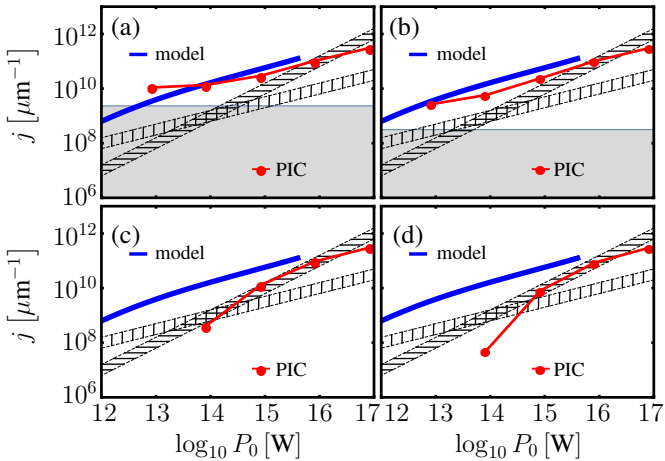} 
\caption{(color online) Comparison of the proton current resulting from a TNSA setup as compared to the model for a lower cutoff energy of (a) $5$ MeV, (b) $10$ MeV, (c) $50$ MeV and (d) $75$ MeV. For comparison: linear scaling (horizontal stripes) and square root scaling (vertical stripes) with the laser power. Gray shaded area: thermal noise.}
\label{fig:TNSA}
\end{figure}
As a benchmark TNSA target we consider a target of heavy ions with a thin proton layer on the backside (s.~Fig.~\ref{fig:Targets} (d)). We find the resulting proton current to closely follow the theoretical predictions in absolute numbers, particularly confirming the expected scaling $j_\text{TNSA}\sim P_0^{1/2}$. Only the data point at $P_0 = 10^{13}$~W does not follow the theoretical prediction (s.~Fig.~\ref{fig:TNSA} (a)). Taking into account that the modeled curve falls below that thermal noise level at this laser power, the observed deviation from the analytical model is most probably due to thermal noise in the simulation. The otherwise good agreement between theory and numerical experiments, however, is by no means surprising, since TNSA has been studied abundantly and is theoretically well understood. Taking into account, on the other hand, that the simulated proton currents at higher lower cutoff energies fall significantly off even for high laser powers confirms that TNSA produces mostly low-energy protons (s.~Fig.~\ref{fig:TNSA} (b)-(d)). These are accelerated, albeit, at a large number, yielding the resultant high proton currents.
\subsection{Comparing high intensity acceleration schemes}
\begin{figure}[t]
\includegraphics[width=\linewidth]{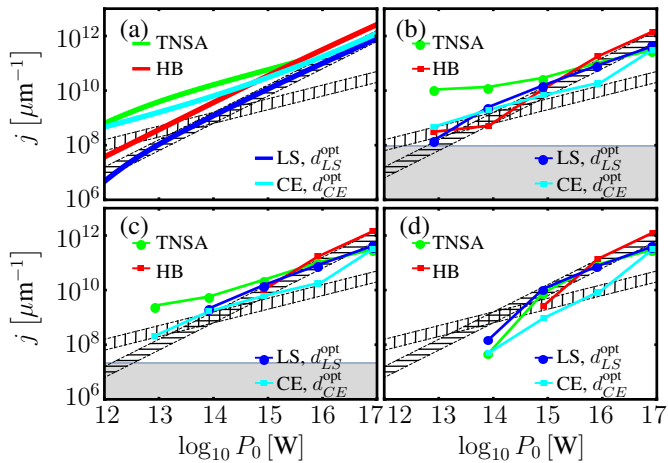}
\caption{(color online) (a): Comparison of the theoretically predicted proton currents resulting from all four studied acceleration setups. [(c)-(d)]: Comparison of the simulated proton currents resulting from all four studied acceleration schemes for a lower cutoff energy of (b) $5$ MeV, (c) $10$ MeV and (d) $75$ MeV. For comparison: linear scaling (horizontal stripes) and square root scaling (vertical stripes) with the laser power. Gray shaded area: thermal noise.} 
\label{fig:PIC_comp}
\end{figure}

Comparing the results from simulations of all the considered schemes (s.~Fig.~\ref{fig:PIC_comp}) one can draw several conclusions. The theoretical models (s.~Fig.~\ref{fig:PIC_comp} (a)) predict TNSA to provide the largest proton currents for laser powers $P_0 \lesssim 10^{15}$~W while at larger powers hole boring provides larger currents, due to its favorable scaling with the laser power. The light sail mechanism is predicted to always yield smaller currents than hole boring, due to the smaller number of accelerated protons, but eventually surpasses the TNSA currents for $P_0 \gtrsim 10^{16}$~W.

The results of the performed two-dimensional PIC simulations largely confirm these predictions. In general, the simulated currents of all protons with energies above $5$~MeV (s.~Fig.~\ref{fig:PIC_comp} (b)) is strongly reminiscent of the theoretical prediction (s.~Fig.~\ref{fig:PIC_comp} (a)). In particular, for laser powers $P_0 < 10^{15}$~W TNSA gives much larger currents than the other schemes, but the current is mostly formed by low-energy protons. When accounting only for protons with energy above $75$~MeV, TNSA is less efficient than light sail and for small laser powers it is also less efficient than Coulomb explosion (s.~Fig.~\ref{fig:PIC_comp} (d)). This is due to the TNSA mechanism quickly establishing strong accelerating fields for a large number of particles. However, the acceleration is spatially confined to the vicinity of the interface between the vacuum and the thermally expanding plasma. The rapid expansion of plasma leads to a quick reduction of the number of protons that experience the accelerating field at this interface. Thus, the process favors acceleration of a large number of protons but to a limited energy. In comparison, Coulomb explosion and light sail provide a control of the number of accelerated protons by tuning the thickness of the target. Thus one can optimize the thickness so that the laser pulse energy is not entirely consumed for accelerating a large number of particles, but more directed towards the acceleration of high-energy protons. Furthermore, light sail is particularly designed for accelerating a limited number of particles to a high energy. As one can see, for laser powers $P_0 \geq 10^{15}$~W the current is almost insensitive to introducing a lower proton energy threshold of up to $75$~MeV (s.~Fig.~\ref{fig:PIC_comp} (b)-(d)), indicating that almost all accelerated have larger energies.

At the range $P_0 < 10^{15}$~W hole boring is difficult to distinguish from TNSA, thus it has similar features, but is substantially less efficient than TNSA. On the other hand, for a laser power above order $P_0 \gtrsim 10^{15}$~W hole boring provides a significantly larger current than all other schemes (s.~Fig.~\ref{fig:PIC_comp} (b)-(d)). This is explained by the fact that it can provide acceleration to a certain level of energy for a large number of protons. This number is proportional to the pulse energy, because the mechanism gives roughly the same acceleration for an increasing number of protons as the pulse penetrates through the target until it is depleted. Thus, if one aims at accelerating a large number of particles to a certain energy in the range of tens of MeV, hole boring seems to provide the optimal strategy for high pulse powers. The scaling law for the hole boring fits well the trend of $j_\text{HB}\sim P_0$. At high powers this establishes its dominance over the TNSA, which follows the trend of $P_0^{1/2}$. 

The light sail mechanism is efficient in terms of delivering the laser power to a limited number of protons, giving them very high energy. In particular, when accounting for only the protons of above $75$~MeV energy, it provides the best results for $P_0 = 10^{15}$~W (s.~Fig.~\ref{fig:PIC_comp} (d)). For larger laser powers, however, the proton current increases less favorably as compared to the hole boring regime, due to a slower rate of growth in the number of accelerated protons.
%
\section{Discussion and outlook}
The ion acceleration mechanisms studied in this work are likely to become the backbone of high-power laser ion-acceleration. Benchmarking them to the broadly employed scheme of ion acceleration, TNSA, we have revealed several promising behaviors and identified strong points as well weaknesses of the theoretical models customarily used to predict the high-power ion acceleration schemes' performance. We have provided a systematic access to the opportunities posed for laser-ion acceleration by currently available as well as future facilities. We introduced a single parameter to quantify a laser-ion accelerator's efficiency in terms of transforming laser energy into accelerated ions. Computing this parameter for various theoretical models of laser-ion acceleration and benchmarking it against the results of elaborate two-dimensional PIC simulations revealed which theoretical models capture the simulated physics reasonably well and which need to be interpreted with care. Furthermore, it showed that for laser powers above $1$~PW essentially all studied high-power ion acceleration schemes surpass the ion currents of the hitherto mostly employed TNSA scheme, especially when taking into account how the latter's current is formed by mostly low-energy ions.

The results presented here can serve as a guidance for the design and planning of upcoming laser-ion acceleration facilities. Comparing the operation parameters of any high-power laser facility to those studied here will allow to assess the accessibility of all studied major laser-ion acceleration schemes as well as the ion currents to be reachable. 

\acknowledgments

This research was supported by the Knut and Alice Wallenberg project PLIONA. The simulations were performed on resources provided by the Swedish National Infrastructure for Computing (SNIC) at PDC and HPC2N.


\end{document}